\documentclass{aa}  

\usepackage{graphicx}
\usepackage{txfonts}
\usepackage{xcolor}

\newcommand{\bl}[1]{\mbox{\boldmath$ #1 $}}

\begin{document}

   
    \title{Gravitoviscous protoplanetary disks with a dust component. II. Spatial distribution and growth of dust in a clumpy disk}
   
\author{Eduard I. Vorobyov\inst{1,2} and Vardan G. Elbakyan\inst{2,3}}
   
\institute{ 
    University of Vienna, Department of Astrophysics, Vienna, 1180, Austria \\
    \email{eduard.vorobiev@univie.ac.at} 
    \and
    Research Institute of Physics, Southern Federal University, Rostov-on-Don, 344090 Russia
    \and 
    Lund Observatory, Department of Astronomy and Theoretical Physics, Lund University, Box 43, 22100 Lund, Sweden
    }

   \date{Received September 15, 2018; accepted March 16, 2018}
   
\titlerunning{Dust distribution and growth in a clumpy disk}

 
  \abstract
   {}
   {Spatial distribution and growth of dust in a clumpy protoplanetary disk subject to vigorous gravitational instability and fragmentation is studied numerically with sub-au resolution using the FEOSAD code.}
   {Hydrodynamics equations describing the evolution of self-gravitating and viscous protoplanetary disks in the thin-disk limit were modified to include a dust
   component consisting of two parts: sub-micron-sized dust and grown dust with a variable maximum radius.  The conversion of small to grown dust, dust growth, friction of dust with gas, and dust self-gravity were also considered.}
   {We found that the disk appearance is notably time-variable with spiral arms, dusty rings, and clumps, constantly forming, evolving, and decaying. As a consequence, the total dust-to-gas mass ratio is highly non-homogeneous throughout the disk extent, showing order-of-magnitude local deviations from the canonical 1:100 value. Gravitationally bound clumps  formed  through  gravitational  fragmentation have a velocity pattern that deviates notably from the Keplerian rotation. Small dust is efficiently converted into grown dust in the clump interiors, reaching a maximum radius of several decimeters. Concurrently, grown dust drifts towards the clump center forming a massive compact central condensation (70--100~$M_\oplus$). We argue that protoplanets may form in the interiors of inward-migrating clumps before they disperse through the action of tidal torques. We foresee the formation of protoplanets at orbital distances of several tens of au with initial masses of gas and dust in the protoplanetary seed in the (0.25--1.6)~$M_{\rm Jup}$ and (1.0--5.5)~$M_\oplus$ limits, respectively. The final masses of gas and dust in the protoplanets may however be much higher due to accretion from surrounding massive metal-rich disks/envelopes.}
   { Dusty rings formed through tidal dispersal of inward-migrating clumps may have a connection to ring-like structures found in youngest and massive protoplanetary disks. Numerical disk models with a dust component that can follow the evolution of gravitationally bound clumps through their collapse phase to the formation of protoplanets are needed to make firm conclusions on the characteristics of planets forming through gravitational fragmentation.}


   \maketitle
%

\section{Introduction}
The early evolution of a protostellar disk can be accompanied by a phenomenon known as gravitational instability \citep[GI, see a review by][]{2016Kratter}, manifesting itself through the development of transient spiral arms in the disk. The susceptibility of protostellar disks to GI has been inferred not only from analytic studies and numerical hydrodynamics simulations \citep[e.g.][]{2007MayerLufkin, 2009StamatellosWhitworth}, but also from recent imaging of spiral arms and arcs in young protostellar disks \citep[e.g.][]{2016Perez,2016LiuTakami}. In the most extreme cases, GI can lead to disk fragmentation if the disk mass is sufficiently high, usually greater than 10\% that of the protostellar mass \citep[e.g.][]{2013Vorobyov}. Such massive disks are rare in the T Tauri phases, but can be ubiquitous in the early embedded phase where disk mass loss via accretion on the protostar, disk winds, and outflows is compensated by continuing mass-loading from the parental collapsing core \citep[e.g.,][]{2010Vorobyov,2011Enoch,2014DunhamVorobyov, 2018Cieza,2018Bate}. The first manifestation of disk fragmentation has recently been observed in L1448 IRS3B by \citet{2016Tobin}.

Disk fragmentation can bring about a number of interesting phenomena, such as FU-Orionis-type outbursts \citep[e.g.][]{2005VorobyovBasu,VorobyovBasu2015,2017MeyerVorobyov,2018ZhaoCaselli} or formation of freely-flowing sub-stellar objects \citep[e.g.][]{2009StamatellosWhitworth,2012BasuVorobyov,2016Vorobyov}, but
perhaps the most interesting and intriguing outcome of disk fragmentation is the possibility of giant planet formation  \citep[e.g.][]{2009Boley,2010Nayakshinb,2015Stamatellos,2013Vorobyov,2013ForganRice,2014GalvagniMayer,2018VorobyovElbakyan}. One problem with the disk fragmentation scenario is that the mass of newly formed gaseous clumps, though initially being in the planetary mass regime, tends to grow via accretion from the surrounding circumstellar disk, thus gradually shifting into the brown dwarf mass regime. This growth can, however, be halted by the radiation output of the forming protoplanet and dynamical scattering \citep[e.g.][]{2017MercerStamatellos} or by the mechanism known as tidal downsizing when the mass of the clump is significantly reduced via stellar torques as it approaches the star  \citep{2010Nayakshinb,2017Nayakshin,2018VorobyovElbakyan}. 

Yet another problem with giant planet formation through disk fragmentation is the ability of gaseous clumps to accumulate dust in their interiors. Observations indicate that giant planets in our Solar system and also giant exoplanets may have substantial solid cores in their interiors.  It is known from theory and numerical simulations that pressure maxima in protoplanetary disks present natural dust traps and can act as sites of dust accumulation and growth, thus assisting the formation of solid protoplanetary cores \citep[see a review by ][]{2016Birnstiel}. In young unstable protostellar disks spiral arms and gaseous clumps can play a role of such dust traps \citep{2004RiceLodato,2008Helled,2010Nayakshinb,2010Nayakshinc,2010BoleyHayfield,2016Booth}. The efficiency of dust trapping in spiral arms, however, is counteracted by the transient nature of spiral arms in protostellar disks and is only efficient near corotation of the spiral pattern with the local gas flow 
\citep{2018VorobyovAkimkin}.  Dust accumulation in gaseous clumps was studied in a series of papers by Nayakshin and co-authors \citep[see a review in][]{2017Nayakshin}, who found that dust can efficiently grow and sediment in the interiors of the clumps. 

In our recent work \citep{2018VorobyovElbakyan}, we studied in detail the inward migration of gaseous clumps formed via disk gravitational fragmentation in the outer disk regions. We demonstrated that gaseous clumps are often perturbed by other 
clumps or disk structures, such as spiral arms, and migrate toward the central star on timescales from a few $10^3$ to few $10^4$~yr. When approaching the star, the clumps quickly lose most of their
diffuse envelopes  through tidal torques. The tidal mass loss helps the clumps to significantly slow down or even halt their inward migration at a distance of a few tens of au from the protostar, 
which can potentially explain the formation of giant protoplanets on tens-of-au orbits in systems such as HR~8799.

In this work, we focus on the efficiency of dust accumulation and growth in a strongly gravitationally unstable and clumpy protostellar disk. For this purpose, we employ the FEOSAD numerical hydrodynamics code developed to study disk formation and long-term evolution taking the dust component into account. Unlike many studies that assume a fixed size of dust grains, we consider dust growth when studying the efficiency of dust accumulation in the disk. The paper is organized as follows. In Sect.~\ref{diskmodel} the numerical hydrodynamics models is briefly reviewed. The main results are presented in Sect.~\ref{results} and summarized in Sect.~\ref{conclusions}.

\begin{table*}
\center
\caption{\label{tab:1}Model parameters}
\begin{tabular}{ccccccccc}
 &  &  &  &  &  &  &  &     \tabularnewline
\hline 
\hline 
Model & $M_{\mathrm{core}}$ & $\beta$ & $T_{\mathrm{init}}$ & $\Omega_{0}$ & $r_{0}$ & $\Sigma_{g,0}$ & $r_{\mathrm{out}}$ & $A$ \tabularnewline
 & {[}$M_{\odot}${]} & {[}\%{]} & {[}K{]} & {[}$\mathrm{km\,s^{-1}\,pc^{-1}}${]} & {[}au{]} & {[}$\mathrm{g\,cm^{-2}}${]} & {[}pc{]} \tabularnewline
\hline 
1 & 1.32 & 0.88 & 20 & 3.1 & 1543 & $1.54\times10^{-1}$ & 0.045 & 1.1 \tabularnewline
2 & 0.99 & 0.77 & 15 & 2.6 & 1474 & $1.26\times10^{-1}$ & 0.045 & 1.2 \tabularnewline
\hline 
\end{tabular}
\center{ \textbf{Notes.} $M_{\mathrm{core}}$ is the initial core
mass, $\beta$ is the ratio of rotational to gravitational energy, $T_{\mathrm{init}}$ is the
initial gas temperature, $\Omega_{0}$ and $\Sigma_{\rm g,0}$ are the angular velocity
and gas surface density at the center of the core, $r_{0}$ is the radius
of the central plateau in the initial core,  $r_{\mathrm{out}}$ is the initial radius of the
core, and $A$ is th einitial positive density perturbation.}
\end{table*}

\section{Protoplanetary disk model}
\label{diskmodel}

When studying numerically young circumstellar disks that are subject to GI and fragmentation the numerical resolution becomes an important factor. This is particularly true when the internal structure of the clumps, which form and evolve in the disk, is in the focus of investigation. The size of the clumps may vary from several au to several tens of au, so that the numerical resolution of several au, a typically allowed minimum in three-dimensional simulations, is not sufficient. Low numerical resolution may cause a pre-mature destruction of the clumps through tidal torques as the clumps migrate inwards, leading to false conclusions on their longevity and survivability.  

To achieve a sub-au resolution, we employ the thin-disk approximation, in which 
negligible vertical motions and the local hydrostatic equilibrium are assumed.
An additional assumption of a small vertical disk scale height with respect to the radial position in the disk (not exceeding 10\%--20\% of the radial distance from the
star in the disk regions of interest as was demonstrated in \citet{2010VorobyovBasu}) allows for the integration of the main hydrodynamic quantities in the vertical direction and the use of these integrated quantities in the hydrodynamics equations.
The obvious advantage of the two-dimensional thin-disk numerical hydrodynamics models is that they can self-consistently describe
the effects of GI and fragmentation (e.g., the formation of spiral arms, and clumps), allowing at the same time for a much better numerical resolution and being computationally inexpensive  in comparison to the full three-dimensional approach. Although intrinsically limited in its ability to model the
full set of physical phenomena that may occur in star and
planet formation, the thin-disk models nevertheless present
an indispensable tool for studying the fine structure of the clumps that form in gravitationally unstable protoplanetary disks. 

The numerical model for the formation and evolution of a star and its circumstellar disk (FEOSAD) is described in detail in \citet{2018VorobyovAkimkin}. Here, we briefly review its main constituent parts and equations.

\subsection{Gaseous disk}
\label{gaseous}
The main physical processes taken into account when modeling the formation and evolution of a gaseous disk include viscous and shock heating, irradiation by the forming star,  background irradiation with a uniform temperature set equal to the initial temperature of the natal cloud core,
radiative cooling from the disk surface, friction with the dust component, and self-gravity of the gaseous and dusty disk. The code is complemented with a calculation of the gas vertical  scale height $H_{\rm g}$ \citep{VorobyovBasu2009}, which is used in the calculation of the fraction of stellar radiation intercepted by the disk and also in the calculation of kinematic viscosity.
The pertinent equations of mass, momentum, and energy transport for the gas component are
\begin{equation}
\label{cont}
\frac{{\partial \Sigma_{\rm g} }}{{\partial t}}   + \nabla_p  \cdot 
\left( \Sigma_{\rm g} \bl{v}_p \right) =0,  
\end{equation}
\begin{eqnarray}
\label{mom}
\frac{\partial}{\partial t} \left( \Sigma_{\rm g} \bl{v}_p \right) +  [\nabla \cdot \left( \Sigma_{\rm
g} \bl{v}_p \otimes \bl{v}_p \right)]_p & =&   - \nabla_p {\cal P}  + \Sigma_{\rm g} \, \bl{g}_p + \nonumber
\\ 
&+& (\nabla \cdot \mathbf{\Pi})_p,
\end{eqnarray}
\begin{equation}
\frac{\partial e}{\partial t} +\nabla_p \cdot \left( e \bl{v}_p \right) = -{\cal P} 
(\nabla_p \cdot \bl{v}_{p}) -\Lambda +\Gamma + 
\left(\nabla \bl{v}\right)_{pp^\prime}:\Pi_{pp^\prime}, 
\label{energ}
\end{equation}
where subscripts $p$ and $p^\prime$ refer to the planar components
$(r,\phi)$  in polar coordinates, $\Sigma_{\rm g}$ is the gas mass
surface density,  $e$ is the internal energy per surface area,  ${\cal P}$
is the vertically integrated gas pressure calculated via the ideal  equation of state as ${\cal P}=(\gamma-1) e$ with $\gamma=5/3$, $\bl{v}_{p}=v_r
\hat{\bl r}+ v_\phi \hat{\bl \phi}$  is the gas velocity in the disk plane, and is $\nabla_p=\hat{\bl r} \partial / \partial r + \hat{\bl
\phi} r^{-1} \partial / \partial \phi $ the gradient along the planar coordinates of the disk.  The gravitational acceleration in the disk
plane,  $\bl{g}_{p}=g_r \hat{\bl r} +g_\phi \hat{\bl \phi}$, takes into account self-gravity of the gaseous and dusty disks found by solving for the Poisson integral \citep[see details in][]{2010VorobyovBasu} and the
gravity of the central protostar when formed. Turbulent viscosity is
taken into account via the viscous stress tensor  $\mathbf{\Pi}$, the expression for which can be found in \citet{2010VorobyovBasu}. We parameterized the
magnitude of kinematic viscosity $\nu=\alpha c_{\rm s} H_{\rm g}$  using the alpha prescription of \citet{1973ShakuraSunyaev} with a constant
$\alpha=10^{-2}$-parameter, where $c_{\rm s}$ is the sound speed of gas. The expressions for the cooling and heating rates $\Lambda$ and $\Gamma$
can be found in \citet{2018VorobyovAkimkin}.

\subsection{Dusty disk}
\label{dustycomp}
In our model, dust consists of two components: small sub-micron-sized dust with a spectrum of radii in the $5\times10^{-3}$--1.0~$\mu$m limits and grown dust with a minimum radius of 1.0~$\mu$m and a variable upper radius $a_{\rm r}$. The slope of the dust size distribution is assumed to have a power-law with an exponent of $p=-3.5$. Initially, all dust is in the form of small sub-micron grains, which constitute the initial reservoir of dust mass in the parental pre-stellar core. As the collapse ensues and disk forms, small dust turns into grown dust, and this process is considered by calculating the dust growth rate and the maximum radius of grown dust. 
The small-to-grown dust conversion rate is defined as
\begin{equation}
\label{GrowthRate}
S(a_{\rm r}) = - {1 \over \Delta t } \Sigma_{\rm d,tot}^n  
{ \int \limits_{a_{\rm r}^n} \limits^{a_{\rm r}^{n+1}} a^{3+p} da \int \limits_{a_{\rm min}} 
\limits^{a_\ast} a^{3+p} da \over \int \limits_{a_{\rm min}} \limits^{a_{\rm r}^n} a^{3+p} da 
\int \limits_{a_{\rm min}} \limits^{a_{\rm r}^{n+1}} a^{3+p} da   },
\end{equation}
where  $\Sigma_{\rm d,tot}=\Sigma_{\rm d,gr}+
\Sigma_{\rm d,sm}$ is the total surface density of dust, $\Sigma_{\rm d,sm}$ and $\Sigma_{\rm d,gr}$ the surface
densities of small and grown dust, respectively, indices $a_{\rm r}^n$ and $a_{\rm
r}^{n+1}$ the maximum dust radii at the current and next time
steps, $a_{\rm min}=0.005~\mu m$ the minimum radius of small dust grains,  and
$a_\ast=1.0$~$\mu$m a threshold value between small and grown
dust components.
The evolution of the maximum radius $a_{\rm r}$ is described as
\begin{equation}
{\partial a_{\rm r} \over \partial t} + (u_{\rm p} \cdot \nabla_p ) a_{\rm r} = \cal{D},
\label{dustA}
\end{equation}
where the growth rate $\cal{D}$ accounts for the dust evolution due to
coagulation and fragmentation. More details on the dust growth scheme can be found in \citet{2018VorobyovAkimkin}.

Small dust is assumed to be coupled to gas, meaning that we only solve the continuity equation for small dust grains, while the dynamics of grown dust is controlled by friction with the gas component and by the total gravitational potential of the star, gaseous and dusty components. The resulting continuity and momentum equations for small and grown dust are
\begin{equation}
\label{contDsmall}
\frac{{\partial \Sigma_{\rm d,sm} }}{{\partial t}}  + \nabla_p  \cdot 
\left( \Sigma_{\rm d,sm} \bl{v}_p \right) = - S(a_{\rm r}),  
\end{equation}
\begin{equation}
\label{contDlarge}
\frac{{\partial \Sigma_{\rm d,gr} }}{{\partial t}}  + \nabla_p  \cdot 
\left( \Sigma_{\rm d,gr} \, \bl{u}_p \right) = S(a_{\rm r}),  
\end{equation}
\begin{eqnarray}
\label{momDlarge}
\frac{\partial}{\partial t} \left( \Sigma_{\rm d,gr} \, \bl{u}_p \right) +  [\nabla \cdot \left( \Sigma_{\rm
d,gr} \, \bl{u}_p \otimes \bl{u}_p \right)]_p  &=&   \Sigma_{\rm d,gr} \, \bl{g}_p + \nonumber \\
 + \Sigma_{\rm d,gr} \bl{f}_p + S(a_{\rm r}) \bl{v}_p,
\end{eqnarray}
where  $\bl{u}_p$ describes the planar components of the grown dust velocity, $S(a_{\rm r}) $ is the rate of dust growth per unit surface area, the expression for which can be found in \citet{2018VorobyovAkimkin}, $\bl{f}_p$ is the drag force per unit mass acting on dust from gas, and $a_{\rm r}$ is the maximum radius of grown dust. We note that the back reaction of dust on gas was not taken into account in this study, but is planned for follow-up studies based on the numerical scheme laid out in \citet{2018Stoyanovskaya}.

Equations~(\ref{cont})--(\ref{energ}) and 
(\ref{contDsmall})--(\ref{momDlarge}) are solved using
the operator-split solution procedure, in which the solution is
split in the transport and source steps. In the transport
step, the update of hydrodynamic quantities due to advection
is done using the third-order piecewise parabolic interpolation
scheme of \citet{1984Colella}. In the source step, the update of hydrodynamic quantities
due to gravity, viscosity, cooling and heating, and also
friction between gas and dust components is performed.
This step also considers the transformation of small to
grown dust and also the increase in dust radius $a_{\rm r}$ due
to growth. We refer the reader to \citet{2018VorobyovAkimkin} for details on the dust growth scheme and vertical dust settling adopted in this work.

We use the polar coordinates ($r,\phi$) on a two-dimensional
numerical grid with $1024\times1024$ grid zones. The radial grid
is logarithmically spaced, while the azimuthal grid is equispaced.
To avoid too small time steps, we introduce a
“sink cell” at $r_{\rm sc}$=12.0 au and impose a transparent inner
boundary condition so that the matter (gas or dust) is allowed
to flow freely from the active domain to the sink cell
and vice versa \citep[for detail see][]{2018VorobyovAkimkin}.
The use of the logarithmically spaced grid in the $r$-direction and equidistant grid 
in the $\phi$-direction allowed us to resolve the disk in the vicinity of the sink
cell with a numerical resolution as small as 0.08 au and achieve a sub-au resolution up to a radial distance of $r=150$~au. The inner 150~au are of particular interest for the current work, because this is where the gaseous clumps form through disk fragmentation, migrate, and disperse. 

\begin{figure*}
\begin{centering}
\includegraphics[width=2\columnwidth]{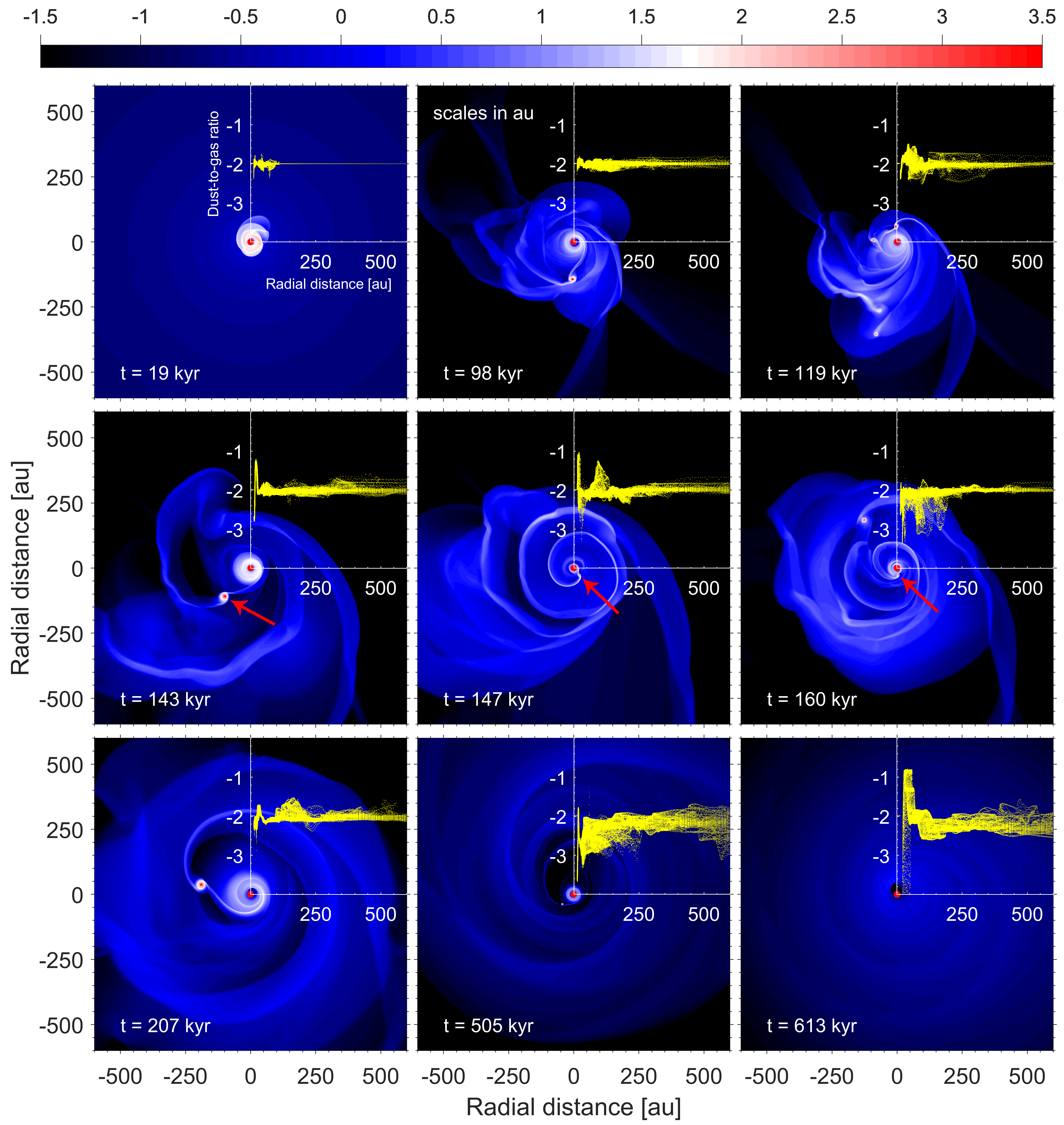}
\par\end{centering}
\caption{\label{fig:9plot_mod2}   Gas surface density maps in model~1 shown for the inner $1200\times1200$~au$^2$ box at nine evolutionary times. The time is counted from the formation of the central star. The disk forms at $t=12.4 \ \rm{kyr}$. The insets in the upper-right corner of each panel present the dust-to-gas mass ratios for all azimuthal grid points at a specific radial distance from the star. The red arrows indicate the position of an inward-migrating clump. The scale bar is in log~g~cm$^{-2}$.}
\end{figure*}

\section{Results}
\label{results}
In this section, we consider the long-term evolution of clumpy protoplanetary disks formed via gravitational collapse of two pre-stellar cloud cores with the initial surface density and angular momentum distributions of the following form: 
\begin{equation}
\Sigma_{\rm g}=\frac{r_{0}\Sigma_{\rm g,0}}{\sqrt{r^{2}+r_{0}^{2}}},
\label{eq:sigma}
\end{equation}
\begin{equation}
\Omega=2\Omega_{0}\left(\frac{r_{0}}{r}\right)^{2}\left[\sqrt{1+\left(\frac{r}{r_{0}}\right)^{2}}-1\right].
\label{eq:omega}
\end{equation}
Cloud cores of these form were shown to form as a result of the slow expulsion of magnetic field due to ambipolar diffusion, with the angular momentum remaining constant during axially-symmetric core compression \citep{1997Basu}. Here,
$\Sigma_{\rm g,0}$ and $\Omega_{0}$ are the
gas surface density and angular velocity at the center of the core, $r_{0}=\sqrt{A}c_{\mathrm{s}}^{2}/\pi G\Sigma_{\rm g,0}$
is the radius of the central plateau, where $c_{\mathrm{s}}$ is the initial isothermal sound speed in the core. In the adopted thin-disk approximation, our pre-stellar cores have the form of
flattened pseudo-disks, a spatial configuration that can
be expected in the presence of rotation and large-scale magnetic fields.

To promote the gravitational collapse of the cores, we impose an initial positive density perturbation \textit{A} set equal to 1.1--1.2, depending on the model. As the collapse proceeds, the inner regions of the core spin up and a centrifugally balanced circumstellar disk forms 
when the inner infalling layers of the core hit the
centrifugal barrier near the sink cell. The material that
has passed to the sink cell before the instance of circumstellar disk formation constitutes a seed for the central star, which further grows through accretion from the circumstellar disk.
The infalling core continues to land at the outer edge of the circumstellar disk until the core depletes. The infall rates on the circumstellar disk are in agreement with what can
be expected from the free-fall collapse \citep{2010Vorobyov}.
The initial gas temperature in collapsing cores $T_{\rm init}$ varies in the 15--20~K limits.    The initial dust-to-gas mass ratio is 1:100. All dust initially is in the form of small dust particles. The initial parameters of the cloud cores are summarized in Table~\ref{tab:1}. The initial mass of the pre-stellar cores and their rotational-to-gravitational energy ratios were chosen so that to produce massive disks prone to GI and fragmentation \citep[see fig.1 in][]{2013Vorobyov}.
The chosen models differ in the mass of the core $M_{\rm core}$, in the amount of initial rotation as defined by $\beta$, and in the initial (and background) temperature as defined by $T_{\rm init}$. These parameters are fundamental for the subsequent evolution of the disk; they define the disk's mass, size, and minimum temperature, which are the parameters that in turn influence the strength of gravitational instability and fragmentation in the disk.  We decided to consider two models with reasonable variations in the properties of initial cores to determine the effect of these variations on the dust evolution in fragmenting disks.


The properties of the nascent star (e.g., stellar radius and photospheric luminosity) are calculated
using the stellar evolution tracks derived using the STELLAR evolution code \citep{2008YorkeBodenheimer}. The evolution of the central star and circumstellar disk are interconnected. The star grows according to the mass accretion rate from the disk, computed as the mass passing through the sink cell per unit time. The growing star also feeds back to disk via radiative heating in accordance with its photospheric and accretion luminosities.

\begin{figure*}
\begin{centering}
\includegraphics[width=2\columnwidth]{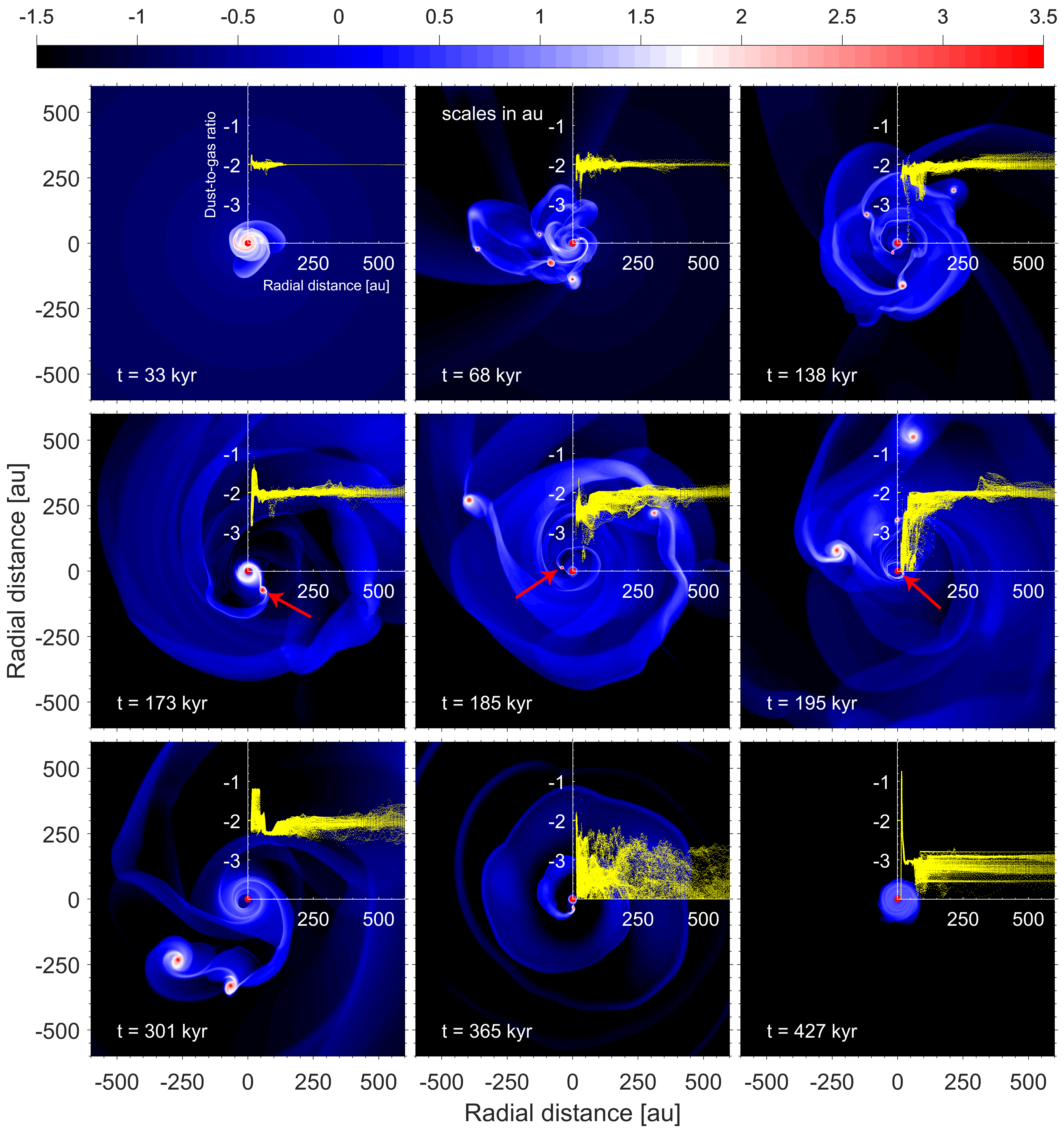}
\par\end{centering}
\caption{\label{fig:9plot_mod3}Similar to Figure \ref{fig:9plot_mod2}, but for model~2.}
\end{figure*}

\subsection{Global evolution of dust in a clumpy protoplanetary disk}

Figure~\ref{fig:9plot_mod2} shows the gas surface density distribution in the inner $1200\times1200$~au$^2$ box of model~1 at nine time instances starting from disk formation and ending after more than 0.6~Myr of disk evolution.  The color scale is chosen so as to highlight the evolving disk sub-structures. The top row demonstrates that the disk rapidly grows in radius from $\approx 50$~au at $t=19$~kyr to several hundreds of au at $t=98$~kyr. In this process, the disk gains enough mass to trigger vigorous GI and fragmentation. Dense gaseous clumps form at radial distances $\ga 50$~au, in agreement with other studies on disk fragmentation \citep[e.g.][]{2009StamatellosWhitworth, 2009Boley}. The red arrows in the middle row indicate one of the clumps that experiences inward migration. We note that at $t=613$~kyr no clumps are present in the disk, implying that their lifetime is not limited by the run time of our numerical simulations, but rather by migration and destruction mechanisms operating in the disk. 

The dynamics of the clumps in the disk, including their radial migration and infall on the star, was studied in detail in our previous works \citep[][see also \citet{2012ZhuHartmann}]{VorobyovBasu2015,2016Vorobyov, 2018VorobyovElbakyan}. In this work, we focus on the dust evolution in a strongly gravitationally unstable disk in general and in clumps in particular. The yellow dots in the insets of Figure~\ref{fig:9plot_mod2} show the radial distribution of the dust-to-gas mass ratio $\zeta_{\rm d2g}=(\Sigma_{\rm d, sm}+ \Sigma_{\rm d, gr})/\Sigma_{\rm g}$ for all azimuthal grid cells at a certain radial distance from the star. Significant deviations of the dust-to-gas ratio from the canonical 1:100 value are evident  and the amplitude of these deviations increases with time. Interestingly, deviations exist both to higher and lower values with respect to 1:100.

In Figure~\ref{fig:9plot_mod3} we present the  gas surface density in the disk of model~2. All notations are similar to Figure~\ref{fig:9plot_mod2}. The disk in model~2 forms at $t=17.2 \ \rm{kyr}$ and its subsequent evolution is qualitatively similar to that of model~1 -- the disk is strongly gravitationally unstable and forms multiple fragments, many of which end up being accreted on the star. The red arrows show one of these dense clumps as it approaches the central star \footnote{The animation of clump inward migration can be found at
http://www.astro.sfedu.ru/animations/model2.mp4}. Similarly to model~1,  the variations in $\zeta_{\rm d2g}$ increase with time and can locally depart from the 1:100 value by more than an order of magnitude.

\begin{figure}
\begin{centering}
\includegraphics[width=1\columnwidth]{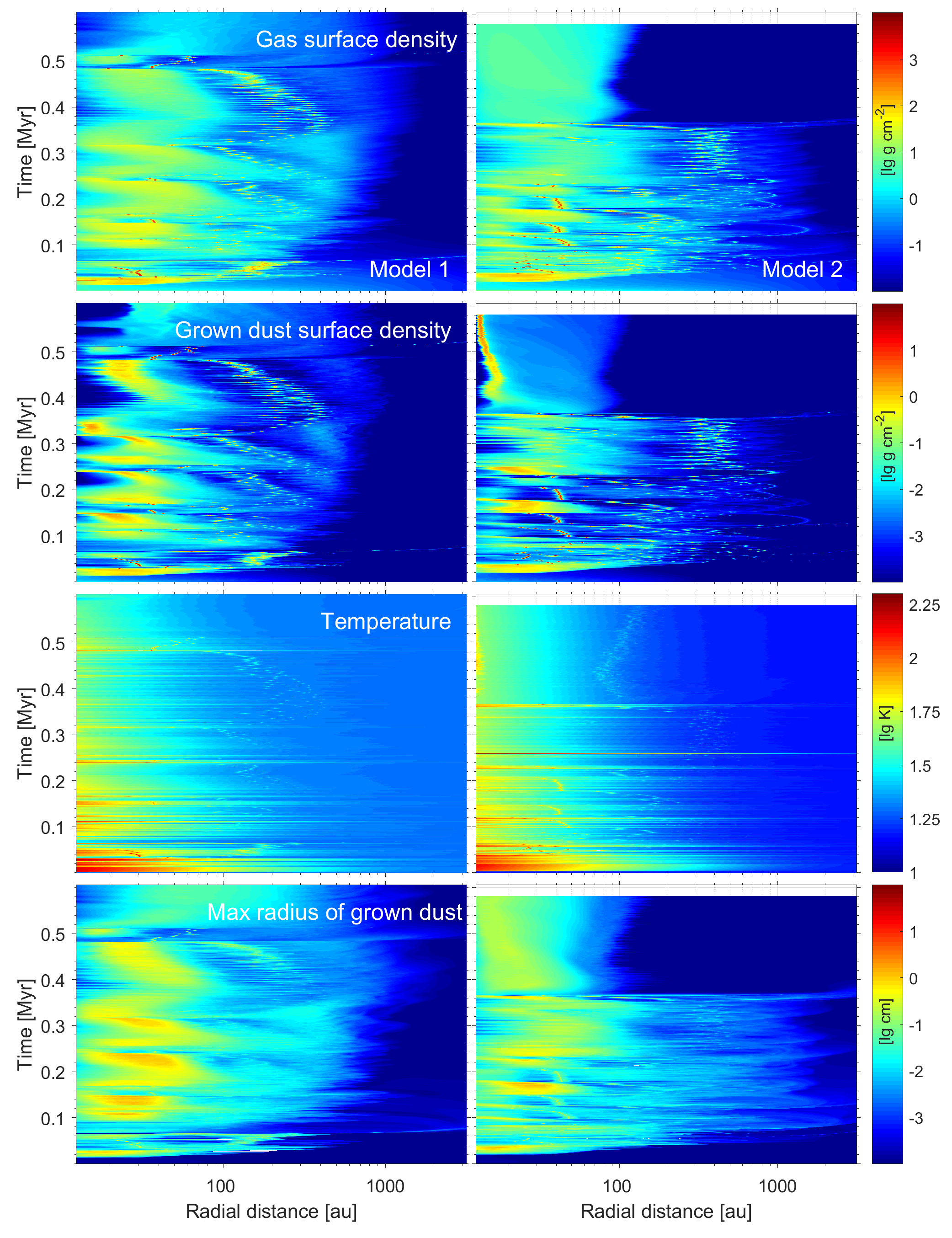}
\par\end{centering}
\caption{\label{fig:12plot}Time-space diagrams showing the temporal evolution of the following azimuthally averaged quantities: surface density of gas (top row), surface density of grown dust (second row), gas temperature (third row), and maximum radius of grown dust (bottom row). The left and right columns correspond to model~1 and model~2, respectively. }
\end{figure}

Figures~\ref{fig:9plot_mod2} and \ref{fig:9plot_mod3} present snapshots of the disk at several discrete evolutionary times. It is also illustrative to show the continuous evolution of the disk starting from its formation till the end of our numerical simulations. In Figure~\ref{fig:12plot}  we present the space-time diagrams showing the temporal evolution of several azimuthally averaged quantities in the disks of model~1 (left column) and model~2 (right column).
The azimuthally averaged gas surface density shown in the top row reveals notable temporal and spatial variations caused by GI and fragmentation. The high surface density speckles correspond to the dense gaseous clumps, which form and migrate in the disk. These panels illustrate the ever changing character of gravitaionally unstable disks with their nonaxisymmetric structures, such as spiral arms and clumps, constantly forming, evolving, and decaying. This tumultuous behaviour of the disk continues for as long as its driving force -- mass-loading from the infalling envelope -- persists.

The second row in Figure~\ref{fig:12plot} presents the temporal evolution of the azimuthally averaged surface density of grown dust, which likewise the surface density of gas demonstrates notable temporal and spatial variations. However, grown dust is characterized by a more compact spatial distribution compared to that of gas, reflecting dust drift and accumulation in local pressure maxima. The high surface density bands represent transient dusty rings, which will be discussed in more detail later in the text.

The third row of Figure~\ref{fig:12plot} presents the temporal evolution of the azimuthally averaged gas temperature. The disk temperature is also highly variable in time and space. Sharp horizontal spikes in the disk temperature reflect mass accretion bursts caused by infall of dense clumps on the protostar \citep[see, e.g.,][]{VorobyovBasu2015}. These accretion events can heat up the inner 100 au of the disk, raising the temperature by more than a factor of 2. Such an increase in the disk temperature can have important consequences for the chemical evolution of the disk \citep[e.g.,][]{2018MolyarovaAkimkin}. 
Finally, the bottom row in Figure~\ref{fig:12plot} presents the azimuthally averaged maximum radius of grown dust, which shows considerable variations throughout both time and space. These variations reflect the non-steady character of evolution of gravitationally unstable disks, which cannot be described by steady-state disk models.

\begin{figure*}
\begin{centering}
\includegraphics[width=\textwidth]{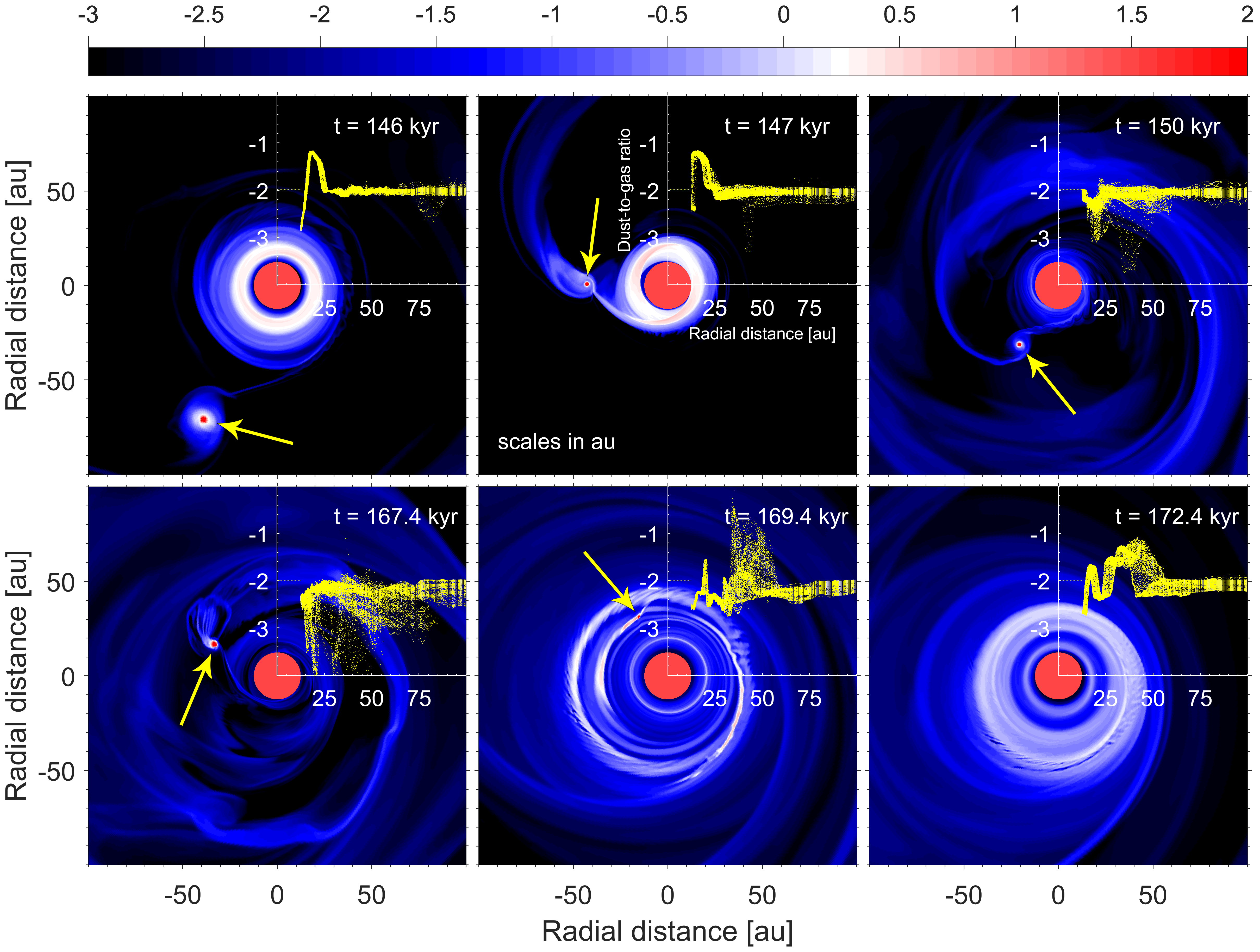}
\par\end{centering}
\caption{\label{fig:ring} Spatial maps of the grown dust surface density in the inner $200\times200$~au$^2$ box shown in model~1 at six consecutive times. The yellow arrows indicate the inward-migrating clump. The insets shows the azimuthal variations in the dust-to-gas ratio at a given radial distance from the star. The red circle in the coordinate center represents schematically the sink cell.}
\end{figure*}

Let us now consider the behaviour of grown dust in Figure~\ref{fig:12plot} in more detail. Clearly, both models show episodic accumulation of grown dust at several tens of au, and these substructures are more pronounced in model~1. To illustrate this phenomenon, we plot in Figure~\ref{fig:ring} the surface density maps of grown dust in the inner $200\times200$ au$^2$ region of model~1. The insets in the top right corner of each panel show the dust-to-gas ratio $\zeta_{\rm d2g}$ for all azimuthal grid cells at a certain radial distance from the star. Initially, two distinct structures are present in the disk: a dusty ring at 20--25~au and a dense clump at 80~au, the latter shown by the yellow arrow. The ring is clearly seen as a local peak in the dust-to-gas ratio (see the insets at $t=146$~kyr and $t=147$~kyr). As the clump migrates inward, it disturbs the ring-shaped dusty structure, forcing part of the dust to accrete through the central sink cell while redistributing the rest on more distant orbits in the disk.

At $t=167.4$ kyr (bottom-left panel) the mass of the clump is 8.7~$M_{\rm Jup}$ including 88.3~$M_{\rm \oplus}$ of dust and it is orbiting at a radial distance of 37~au from the protostar. At $t\approx169.4$~kyr the clump is teared apart by tidal torques, a process studied in detail in \citet{2018VorobyovElbakyan}. As a result, the clump releases its dust content to the disk, forming a ring-shaped dusty structure in the inner several tens of au. At $t=172.4$~kyr the clump is totally disintegrated and the dusty ring with an elevated $\zeta_{\rm d2g}$ is re-formed again.
The total mass of the re-formed dusty ring is about $80~M_{\rm \oplus}$, which is in agreement with the total dust mass  inside the disintegrated clump.

Ring-like structures in the dust continuum emission have been detected for a number of protoplanetary disks with ages ranging from $<0.5$~Myr to $>10$~Myr \citep{2019vanderMarel}. The origin of these structures is debated  and may be caused by different mechanisms, such as embedded planets \citep[e.g.,][]{2015DongZhu}, ice lines \citep{2015Zhang}, and dead zones \citep[e.g.,][]{2014ChatterjeeTan,2015Flock,2019Kadam}. 
The cyclic appearance and disappearance of dusty rings observed in our models takes place in the initial gravitationally-unstable stages of disk evolution. As Figure~\ref{fig:12plot} demonstrates, this stage does not last much longer than 0.5~Myr. In the sample studied by \citet{2019vanderMarel}, only one object, Elias~24, is sufficiently young to fall in this category. However, its disk mass estimated from the dust continuum, $53~M_{\rm Jup}$, may not be sufficient for disk fragmentation to develop, for which process a disk mass in excess of 80--100~$M_{\rm Jup}$ is needed \citep{2013Vorobyov}. 

Two things have to be noted in this context. First, the determination of stellar ages suffers from uncertainties in the details of stellar evolution in the main accretion phase. As was demonstrated by, e.g., \citet{2018VorobyovElbakyan}, variations in the amount of accretion energy absorbed by the star (cold vs. hot accretion) can change significantly the stellar evolution tracks of stars with ages $<$~a~few~Myr, making them look older than they are. If the disk age is overestimated by factors of 2--3, then AS~209, GY~91, HL~Tau, and Sz~98 will be brought into young-disk category.  Second, disk mass estimates from the thermal emission of dust suffers from uncertainties related to the dust temperature, dust-to-gas conversion rate, disk inclination, and optical depth effects \citep[][]{2014DunhamVorobyov}. Errors can be as large as factors of several to ten, and often resulting in disk mass underestimates rather than overestimates, which can shift disk masses to the gravitational unstable regime for at least Elias~24 and HL~Tau. We note that a spiral arm in the line emission of HCO$^+$ has recently been detected for HL~Tau \citep{2019Yen}. We therefore cannot rule out the relation of our dusty rings to at least some ring-like structures detected by ALMA at distances on the order of several tens of au. A detailed comparison of the observed ring-like structures with the synthetic dust continuum images of our disks is needed to make firmer conclusions.


\subsection{Dust dynamics and growth in gaseous clumps}

\begin{figure}
\begin{centering}
\includegraphics[width=1\columnwidth]{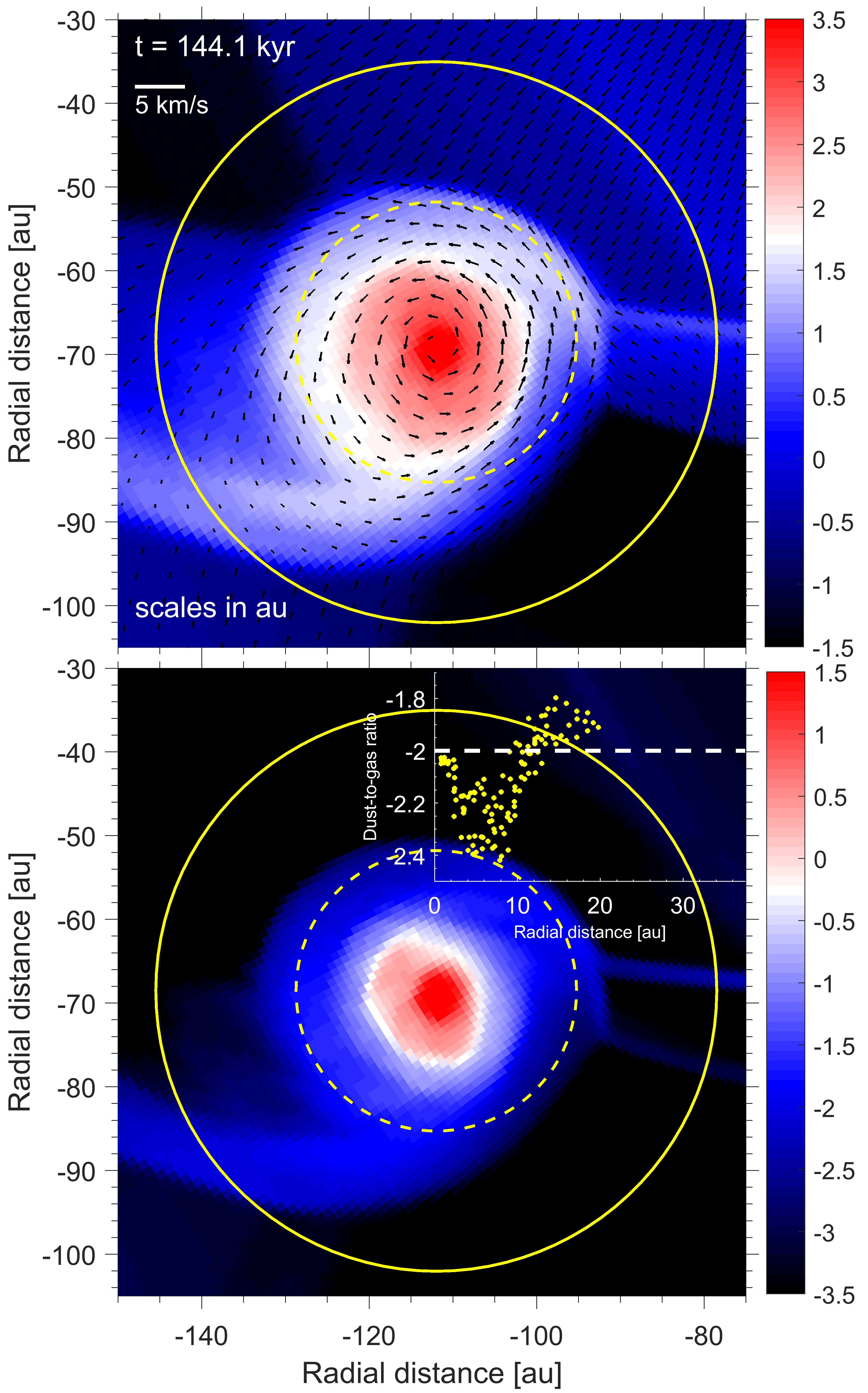}
\par\end{centering}
\caption{\label{fig:1}{\bf Top panel}: Gas surface density distribution (in log g cm$^{-2}$) in the vicinity of the clump at $t=144.1$~kyr in model~1. The superimposed black arrows show the velocity field of the gas.  {\bf Bottom panel}: Grown dust surface density distribution (in log g cm$^{-2}$) in the vicinity of the clump. The inset shows the azimuthal variations of the dust-to-gas ratio (with respect to the clump center) vs. radial distance from the clump center. The dashed white line shows the canonical value of 1:100 for the dust-to-gas ratio.  In both panels the solid yellow circles mark the outer edge of the Hill sphere of the clump. The dashed yellow circles have a half radius of the Hill sphere. }
\end{figure}

In this section we focus on the evolution of individual clumps to obtain a better understanding of the clump internal structure during its inward migration in the disk. For this purpose, we take the clump indicated in Figure~\ref{fig:9plot_mod2} with the red arrows. Figure~\ref{fig:1} presents part of the disk containing the chosen clump and illustrates the gas and dust dynamics in the vicinity of the clump. The coloured maps in the top and bottom panels demonstrate the surface densities of gas and grown dust, respectively. 
The black arrows in the top panel represent the gas velocity field superimposed on the gas surface density distribution.
The yellow solid circles show the size of the Hill radius of the clump defined as
\begin{equation}
    R_{\mathrm{H}} = R_{\mathrm{cl}} \Bigg(\frac{M_{\mathrm{cl}}}{3(M_{\mathrm{*}}+M_{\mathrm{cl}})}\Bigg)^{1/3},
\end{equation}
where $M_{\mathrm{cl}}$ and $M_{\mathrm{*}}$ are the masses of the clump and the central star, respectively,  and $R_{\mathrm{cl}}$ is the radial distance of the clump from the protostar.
Motivated by the study of \citet{2017NayakshinDesert}, who found that the gas bound to the protoplanet is usually located within the inner half of the Hill radius and the material in the outer half of the Hill radius is much more likely to be lost as the protoplanet migrates inwards, we also introduced the yellow dashed circle which has a radius of $R_{\mathrm{H}}/2$. Interestingly, the gas inside the inner half of the Hill radius tends to rotate around the center of the clump on near-circular orbits with velocities up to a few km s$^{-1}$, while the gas in the outer half of the Hill radius shows notable deviations from a circular motion. 
The inset in the bottom panel of Figure~\ref{fig:1} shows the azimuthal variations of the dust-to-gas ratio at a certain radial distance from the center of the clump. The horizontal dashed line marks the canonical value of 1:100. Clearly, the inner region of the clump ($\le 12$~au) is characterized by the highest surface densities of grown dust, while simultaneously having a dust-to-gas ratio that is lower than the 1:100 value. The situation is reversed in the outer regions of the clump. 

\begin{figure}
\begin{centering}
\includegraphics[width=1\columnwidth]{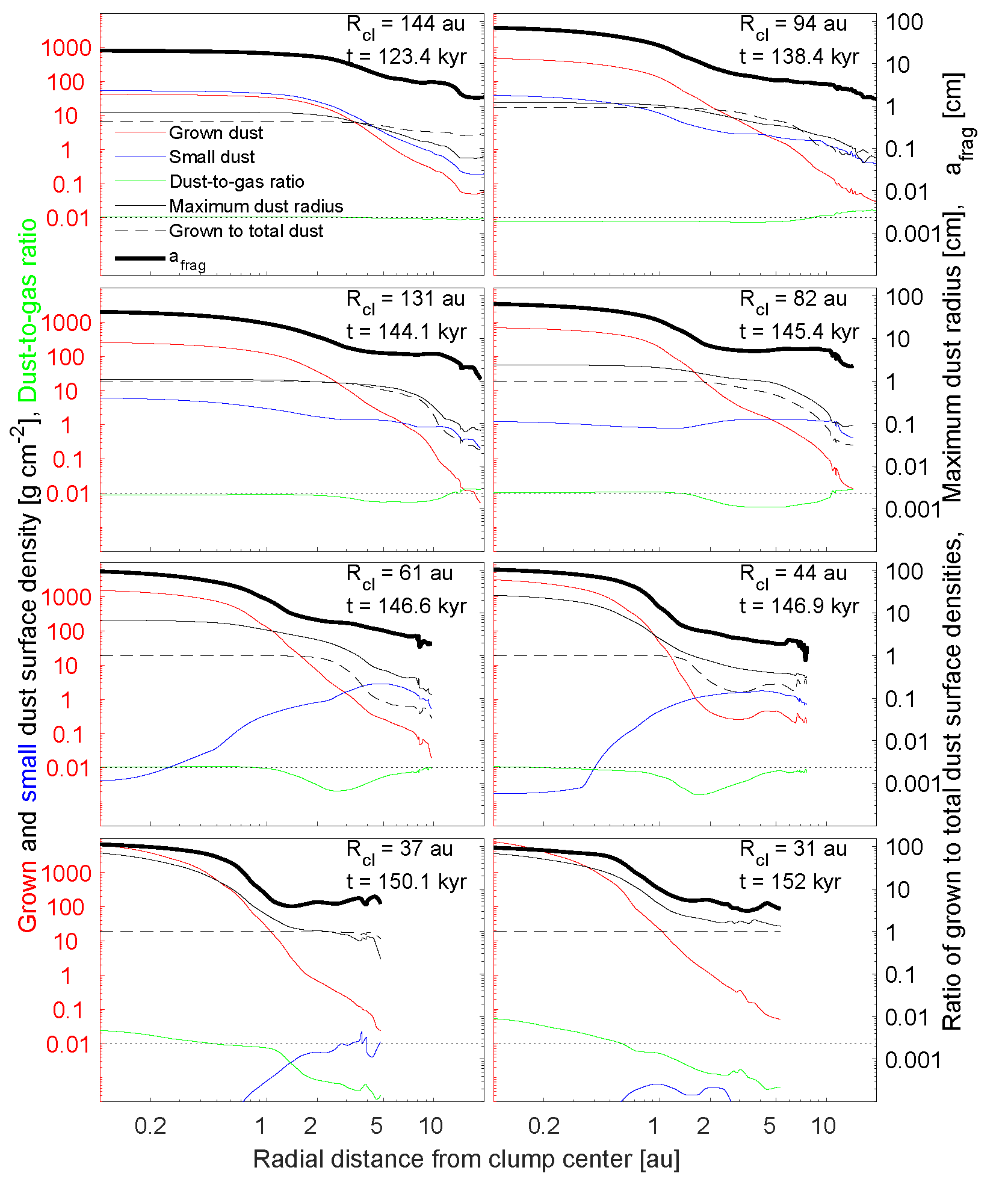}
\par\end{centering}
\caption{\label{fig:2} Azimuthally averaged radial distributions of the grown dust (red lines) and small dust (blue lines) surface densities, total dust to gas ratio (green lines), maximum radius of dust grains (thin black solid lines), ratio of grown to total dust surface densities (dashed black lines), and fragmentation barrier $a_{\rm frag}$ (thick black solid lines) vs. radial distance from the center of the clump. The evolutionary time and the radial distance of the clump from the star $R_{\rm cl}$ are indicated in each panel. The horizontal dotted lines mark the canonical dust-to-gas ratio of 1:100 for convenience. } 
\end{figure}

To study in more detail the radial distribution and time evolution of the dusty component inside the clump as it migrates inward, we calculated the azimuthally averaged (with respect to the center of the clump) characteristics of the clump. Figure~\ref{fig:2} presents the azimuthally averaged radial distribution of grown (red lines) and small (blue lines) dust surface densities, dust-to-gas mass ratio $\zeta_{\rm d2g}$ (green lines), maximum radius of grown dust $a_{\rm r}$ (thin black solid lines), ratio of grown to total dust surface densities (dashed black lines), and fragmentation barrier $a_{\rm frag}$ (thick black solid lines) vs. the radial distance from the center of the clump. Several time instances are shown starting from the clump formation (upper-left panel) and ending near its dispersal (bottom-right panel). We note that the radial distance in Figure~\ref{fig:2} is presented in the log scale, which allows us to show in more detail the inner region of the clump. The dust fragmentation barrier is defined as 
\begin{equation}
\label{afrag}
a_{\rm frag} = \frac{2\Sigma_{\rm g}u^2_{\rm frag}}{3\pi \rho_{\rm s} \alpha c_{\rm s}^2},
\end{equation}
where $u_{\rm frag}$=30~m~s$^{-1}$ is a threshold value for the dust fragmentation velocity, and $\rho_{\rm s}$ = 2.24 g cm$^{-3}$ is the material density of dust grains. 

Initially, small dust dominates throughout the clump extent, but already after 15~kyr of clump evolution ($t=138.4$~kyr) grown dust starts dominating in the inner parts  and after 25~kyr ($t=150.1$~kyr) virtually all small dust is converted in the grown form. Concurrently,
the maximum radius of grown dust increases and approaches the fragmentation barrier.  The efficient conversion of small to grown dust is also evident from the dashed black lines showing the ratio of grown dust to the total dust content -- the ratio approaches unity after 25~kyr of clump evolution throughout the clump extent.  
We note that the transitional radius between small and grown dust in our model is $a_{\rm \ast}=1.0$~$\mu$m, meaning that only sub-micron particles have in fact been depleted. A multi-band model for dust growth is needed to further study the conversion efficiency of small to grown dust grains in gaseous clumps formed through disk fragmentation. 

Notable changes in the dust-to-gas mass ratio $\zeta_{\rm d2g}$ are also evident during the inward migration of the clump. Initially at $t=123.4$~kyr, $\zeta_{\rm d2g}$ in the inner parts of the clump is close to the standard 1:100 value. 
During the subsequent inward migration, the dust-to-gas ratio in the inner parts becomes higher than the 1:100 value up to a factor of 5, while in the rest of the clump $\zeta_{\rm d2g}$ drops below the canonical by up to a factor of ten. The elevated $\zeta_{\rm d2g}$ in the inner regions of the clump and reduced $\zeta_{\rm d2g}$ in the outer regions imply that the grown dust has efficiently drifted inwards. We also emphasize that during the inward migration of the clump its radius has decreased from about 20~au at $t$=144.1~kyr to about 5~au at $t$=152~kyr. Other clumps, e.g., the clump of model~2 indicated in Fig.~\ref{fig:9plot_mod3} with the red arrows, show similar trends.


\begin{figure}
\begin{centering}
\includegraphics[width=1\columnwidth]{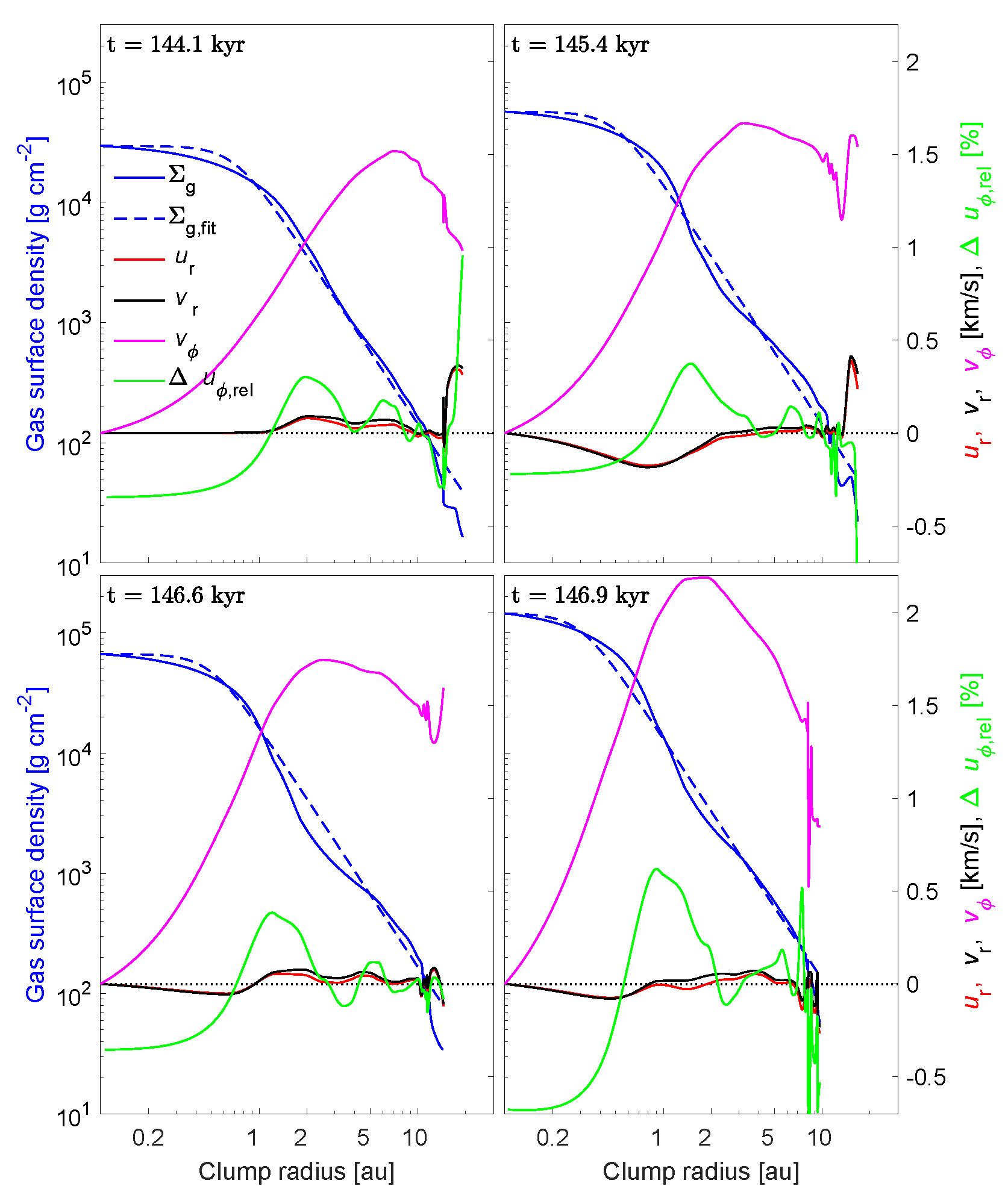}
\par\end{centering}
\caption{\label{fig:9plot_vel} Radial profiles of the gas surface density (blue lines), radial velocity of gas (black lines) and grown dust (red lines), azimuthal velocity of gas (pink lines), along with the relative deviation of the dust velocity from that of gas in per cent (green lines) shown as a function of the radial distance from the center of the clump. The time instances are indicated in each panel. 
The dashed blue line shows the best fit for the gas surface density profile as provided by Equation~(\ref{bestfit}). The horizontal dotted lines mark the zero value of velocity for convenience.}
\end{figure}

To further study the structure and internal dynamics of the clump of model~1, we plot in Figure~\ref{fig:9plot_vel} the azimuthaly averaged (with respect to the center of the clump) radial velocities of gas and grown dust, azimuthal velocity of gas, and gas surface density as a function of radial distance from the center of the clump at several time instances. The gas surface density profile of the clump consists of a central plateau and  a power-law outer tail. The best-fit curves for the gas density profiles are shown with the blue dashed line and have the following form
\begin{equation}
\Sigma_{\rm g}={\Sigma_{\rm cl,0} \over \sqrt{1+\left({r\over r_{\rm cl,0} }\right)^4}},
\label{bestfit}
\end{equation}
where $\Sigma_{\rm c,0}$ is the gas surface density at the center of the clump and $r_{\rm cl,0}$ is the radius of the central plateau. The value of $r_{\rm cl,0}$ is decreasing from 0.7~au at $t=144.1$~kyr to 0.3~au at $t=146.9$~kyr. At the same time the value of $\Sigma_{\rm cl,0}$ is increasing from $3\times10^4$~g~cm$^{-2}$ to $1.5\times10^5$~g~cm$^{-2}$, meaning that the central core becomes denser and more compact.

The azimuthal velocities of gas and grown dust in the clump are very close to each other. Therefore, we show with the magenta lines only the azimuthal velocities of gas $v_{\phi}$. In addition, we plot with the green lines the relative deviation of the azimuthal velocity of grown dust from that of gas in per cent defined as $\triangle u_{\rm \phi,rel}=100 (u_\phi - v_\phi)/v_\phi$. The radial profile of $\triangle u_{\rm \phi,rel}$ has a complicated pattern, but positive values generally dominate over negative ones, implying that grow dust
is overall decelerated via friction with gas when orbiting around the center of the clump. The radial velocities of grown dust and gas also show a complicated pattern. However, one can notice that the inner parts of the clump are generally characterized by negative radial components of velocities, while in the outer parts the corresponding components are predominantly positive. This means that the inner regions of the clump contract, while the outer parts expand and may finally be lost through the Hill sphere via the action of tidal torques.
Finally, we note that the velocity pattern at the clump outer edge shows a rather chaotic behaviour, which can be explained by the perturbing influence of the surrounding circumstellar disk.

Interestingly, the azimuthal velocities of gas and grown dust are significantly different from what can be expected for a near-Keplerian circumstellar disk having by a radially declining azimuthal velocity. On the contrary, the inner regions of the clump are characterized by the azimuthal velocities that increase with distance, which is a consequence of a near-constant density plateau in the center of the clump. For a Keplerian velocity profile to exist, one needs a massive point-sized gravitating body in the center, which is obviously absent at this stage of the clump evolution. Only in the outer parts of the clump the azimuthal velocity starts declining with distance, meaning that the central core now contributes to the gravitational potential at these distances as a near-point-sized object.


\begin{figure}
\begin{centering}
\includegraphics[width=1\columnwidth]{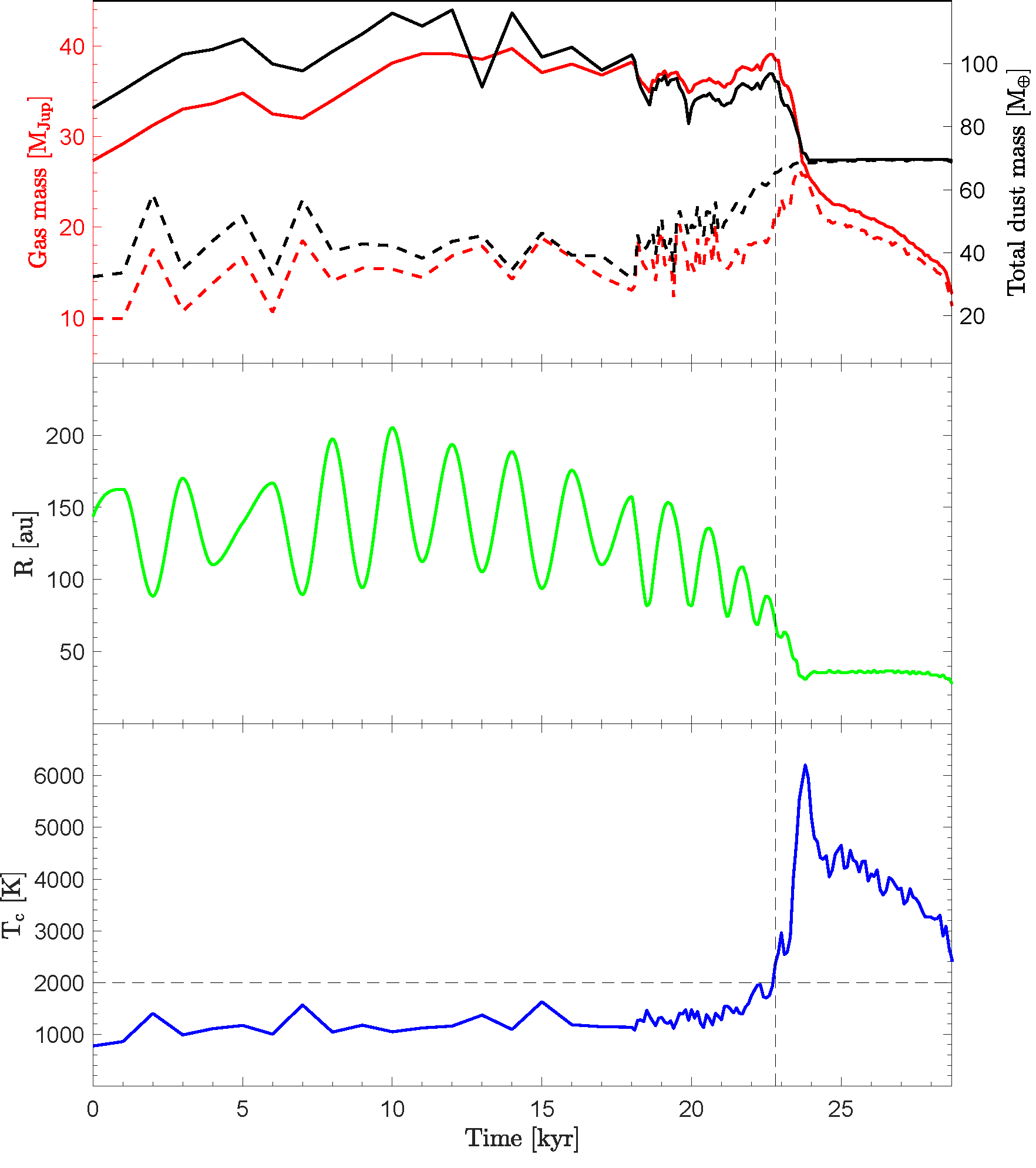}
\par\end{centering}
\caption{\label{fig:3} {\bf Top panel}. Gas and total dust masses (red and black solid lines, respectively) of the clump vs. time. The red and black dashed lines show the gas and total dust mass inside the inner 2~au of the clump, respectively. {\bf Middle panel}. The radial distance of the clump from the central star vs. time. {\bf Bottom panel}. The gas temperature in the center of the clump vs. time. The horizontal dashed line marks a threshold temperature of 2000~K, above which molecular hydrogen dissociates. In all panels the vertical dashed lines show the time instance when the central temperature of the clump reaches the threshold value. The time is counted from $t=123.4$~kyr.
}
\end{figure}

We now focus on the temporal evolution of the main characteristics of the clump as it approaches the star. Figure~\ref{fig:3} presents the gas and total dust masses (top panel), radial distance of the clump from the star (middle panel), and central temperature of the clump (bottom panel) as a function of time. 
More specifically, the solid lines in the top panel show the gas and dust masses of the entire clump, while the dashed lines present the masses contained inside the inner 2~au of the clump.
The time is counted from  the instance of the clump formation in the disk, $t=123.4$~kyr, and is referred below with the Greek letter $\tau$.

The clump forms at a distance of about 150~au and initially shows large-amplitude oscillatory motions as it orbits the star. In this process, the clump gains a peak gas mass of about 42~$M_{\rm Jup}$. The dust mass also increases and reaches a peak value of 100 Earth masses. The temperature in the clump center slowly increases, but only slightly exceeds 1000~K. After about $\tau=10$~kyr the clump starts its migration towards the star, a process often triggered by the gravitational interaction with other clumps in the disk \citep{2018VorobyovElbakyan}, until it settles at a quasi-stable orbit of about 30~au. A sharp mass loss of the clump halts its inward migration, a phenomenon discussed in more detail \citet{2018VorobyovElbakyan}. During the inward migration the gas mass is reduced by a factor of several (from 42~$M_{\rm Jup}$ to 12 $M_{\rm Jup}$). Interestingly, the dust mass does not show such a dramatic decrease -- it drops from 100~$M_\oplus$ to 70~$M_\oplus$. As the clump migrates inwards, it loses its material through the shrinking Hill sphere via the action of tidal torques. A much faster loss of gas means that the clump's outer regions (which are in the first place subject to tidal striping) have been depleted in dust and most dust has already migrated towards the center of the clump -- a process that we have also witnessed in Figure~\ref{fig:2}.    
We also note that while the gas mass inside a 2~au radius continues to decrease in the final stages of the clump evolution, the dust mass stays constant, meaning that the size of the central dusty core is quite compact. 

The clump finally disperses after $\tau=30$~kyr of evolution. In reality, however, the second collapse of the clump to stellar densities should have occurred before the clump dispersal. The horizontal dashed line in the bottom panel shows a threshold temperature of 2000~K, above which molecular hydrogen dissociates and the  clump is supposed to collapse to stellar densities, thus forming a protoplanet, which can withstand tidal torques at these distances from the star. We note that this process is not captured in our numerical simulations because of limited numerical resolution but its consequences are discussed later in Section~\ref{secondcollapse}.



\begin{figure}
\begin{centering}
\includegraphics[width=1\columnwidth]{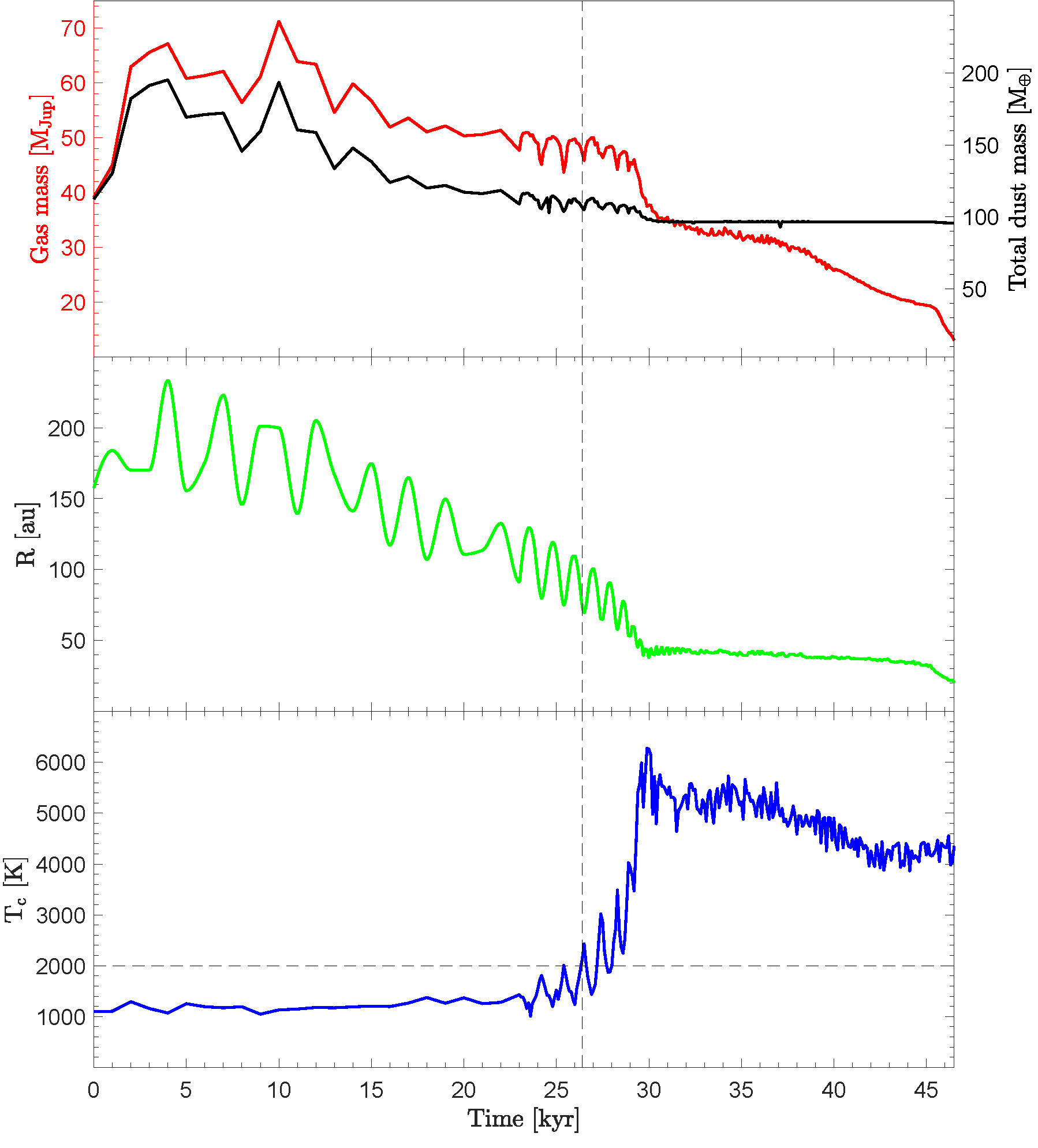}
\par\end{centering}
\caption{\label{fig:7} Similar to Figure \ref{fig:3}, but for the clump indicated in Figure~\ref{fig:9plot_mod3} with the red arrows.}
\end{figure}

To make sure that the above described process is typical of a inward-migrating clump, we consider another clump from model~2 indicated in Figure~\ref{fig:9plot_mod3} with the red arrows. The time evolution of the gas and total dust masses in the clump, its radial distance from the star, and its central temperature are presented in Figure~\ref{fig:7}. The time is counted from the instance of clump formation at $t=152.4$~kyr. The notations in the Figure are the same as in Figure~\ref{fig:3}. The evolution of the clump in model~2 is qualitatively similar to that of model~1. Both clumps lose most of its initial gas mass when migrating inwards, but retain a substantial fraction of the dust mass. Both settle at a quasi-steady orbit at several tens of au, but ultimately disperse because of tidal torques. In both clumps the gas temperature in the center of the clump exceeds a threshold temperature of 2000~K before clump dispersal, implying that a protoplanet could have formed at a certain stage of clump evolution. 
Table~\ref{tab:3} summarizes the total (gas plus dust) masses and total dust (small plus grown) masses of the clumps at the instance of their dispersal. Clearly, the clumps are overabundant in dust as compared to the canonical 1:100 value. 
In the future, we plan to introduce a sub-grid model for protoplanets to study the evolution of dust in their primordial atmospheres in the evolving disk \citep{2018Ragossnig}.

\begin{table}
\center
\caption{\label{tab:3} Characteristics of gaseous clumps at the instance of dispersal}
\begin{tabular}{cc|c}
\hline 
& model 1 & model 2 \tabularnewline
\hline 
\hline 
Total mass [$M_{\rm Jup}$]  & 12.7 & 13 \tabularnewline
Dust mass [$M_{\oplus}$]  & 68.8 & 95.7 \tabularnewline
Dust-to-gas ratio & 0.017 & 0.023 \tabularnewline
\hline 
\end{tabular}
\end{table}


\subsection{The second collapse and prospects for giant planet formation}
\label{secondcollapse}
Although our numerical model does not consider molecular hydrogen dissociation at temperatures exceeding 2000~K and our still limited numerical resolution does not allow us to compute the subsequent formation of protoplanets through the process known as the second collapse \citep[e.g.,][]{2000MasunagaInutsuka}, we nevertheless can estimate the main characteristics of the would-be formed protoplanet using the surface density and angular velocity profiles of the clump at the time instance when the gas temperature in the clump interiors exceeds 2000~K. For this purpose, we consider the clump of model~1 shown in Figure~\ref{fig:9plot_mod2} with the red arrows. 
Just before the second collapse, the clump was characterized by a rather high ratio of rotational to gravitational energy of $\beta_{\rm cl}=25\%$. Therefore, we expect that only a small fraction of the clump material could collapse directly to the protoplanet, while the rest could form a massive disk and/or envelope.

\begin{figure}
\begin{centering}
\includegraphics[width=1\columnwidth]{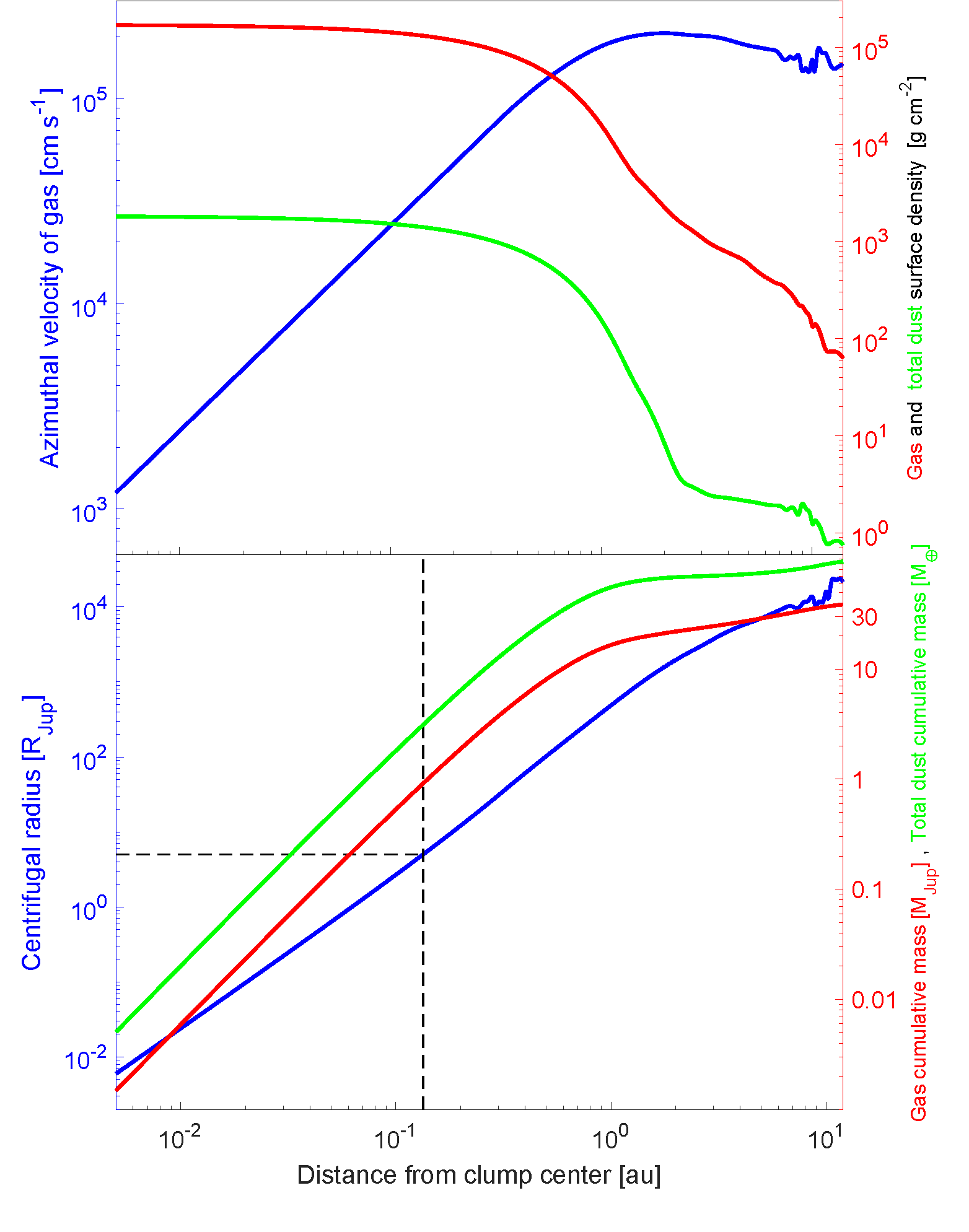}
\par\end{centering}
\caption{\label{fig:8} {\bf Top panel}. Azimuthally averaged surface density of gas (red line) and total dust (green line), along with the angular velocity profile of the clump (blue line), vs. the radial distance from the center of the clump. 
{\bf Bottom panel}. The
centrifugal radius (blue line) and the cumulative gas (red line) and total dust (green line) mass of the clump vs. the radial distance from the center of the clump. The horizontal dashed line marks the adopted radius of the protoplanet $R_{\rm p.p.}=5 R_{\rm Jup}$. The vertical dashed line separates the inner and outer parts of the clump that would form the protoplanet and the disk/envelope, respectively, after the second collapse.}
\end{figure}

The top panel of Figure~\ref{fig:8} shows the azimuthally averaged gas and total dust surface densities (red and green lines, respectively) and angular velocity of gas in the frame of reference of the center of the clump (blue line) vs. the radial distance from the center of the clump. The profiles are shown at $\tau=22.8$~kyr  when the central temperature of the clump reaches a critical value of 2000~K (see Figure~\ref{fig:3}). The red and green lines in the bottom panel of Figure~\ref{fig:8} show the azimuthally averaged cumulative masses of gas and total dust, respectively. We note that the innermost data point for the clump lies at $\approx0.3$~au from its center and the data at smaller radii were derived through extrapolation assuming that the angular velocity $v_{\rm \phi}=0$ at $r=0$ and the surface densities of gas and total dust turn into a constant plateau in the central part of the clump.

To estimate the characteristics of the protoplanet that could  form through the second collapse of the clump, we calculate the centrifugal radii of the material constituting the clump as follows:
\begin{equation}
\label{Rcf}
R_{\rm cf}(r^\prime) = \frac{|\boldsymbol{J}|^2}{GM_{\rm c}(r^\prime)}
\end{equation}
where $M_{\rm c}(r^\prime)$ is the mass contained within the radial distance $r^\prime$ from the center of the clump and $\boldsymbol{J}$ is the specific angular momentum of the clump, which is assumed to be conserved during the second collapse.

The resulting values of $R_{\rm cf}(r^\prime)$  are presented in the bottom panel of Figure~\ref{fig:8} with the blue line. The initial radius of the protoplanet is taken to be equal to $R_{\rm p.p.}=5 R_{\rm Jup}$ (shown with the horizontal dashed line), which is within the limits suggested by various studies \citep[e.g.,][]{2011HosokawaOffner, 2017VorobyovElbakyan}. The vertical dashed line marks a critical radius of $R_{\rm crit}=0.18$~au. Part of the clump that lies to the left of the vertical dashed line would directly form the protoplanetary seed during the second collapse, while the material that lies to the right of the vertical dashed line would hit the centrifugal barrier before accreting on the protoplanet and would therefore form a circumplanetary disk and/or envelope. At $\tau=22.8$ kyr the total mass of the clump is 38.5~$M_{\rm Jup}$ (including the total mass of dust). We found that only a small portion of the clump would eventually form the protoplanetary seed with a mass of 0.91~$M_{\rm Jup}$, including the dusty component with a mass of 3.1~$M_{\oplus}$. 

Using the same approach, we calculated the gas and dust masses of the protoplanetary seed that could have formed after the second collapse of the clump shown in Figure~\ref{fig:9plot_mod3} with the red arrows (model~2). The total mass of the clump before the second collapse is 47.1~$M_{\rm Jup}$. Only 0.48~$M_{\rm Jup}$ would eventually form the protoplanetary seed, including 1.9~$M_{\oplus}$ of dust. The clump in model~2 is characterized by a ratio of rotational to gravitational energy of $\beta=33\%$, which is higher compared to the clump in model~1. Despite a higher mass of the clump in model~2, the resulting protoplanetary seed is less massive than the one of model~1.


To check how our results depend on the chosen value of $R_{\rm p.p.}$,  we also calculated the masses of the protoplanetary seeds  assuming $R_{\rm p.p.}=2.5~R_{\rm Jup}$ and $R_{\rm p.p.}=10~R_{\rm Jup}$. The protoplanetary seed of model~1 with a factor of 2 smaller $R_{\rm p.p.}$ would have a mass of 0.5~$M_{\rm Jup}$, including the dusty component with a mass of 1.7~$M_{\oplus}$, while for $R_{\rm p.p.}=10~R_{\rm Jup}$ the protoplantery mass would be  1.62 $M_{\rm Jup}$, including the dusty component with a mass of 5.51~$M_{\oplus}$. The protoplanetary seed in model~2 would have  a mass of 0.25~$M_{\rm Jup}$, including the dusty component with a mass of 1.0~$M_{\oplus}$ when $R_{\rm p.p.}=2.5~R_{\rm Jup}$ and a mass of 0.87~$M_{\rm Jup}$, including the dusty component with a mass of 3.5~$M_{\oplus}$ when $R_{\rm p.p.}=10~R_{\rm Jup}$. The results are summarized in Table~\ref{tab:2}. Clearly, the protoplanetary seeds are slightly overabundant in dust as compared to the  1:100 value. 

\subsection{Model caveats}

Several caveats need to be mentioned when considering our results. First, the masses of the dusty component inside the protoplanetary seeds were calculated not taking dust sublimation into account. Part of the dust may sublimate in the hottest innermost regions of the clump before the second collapse ensues. Whether these heavy elements are retained in the primordial atmosphere around the protoplanet or rain back on its solid core remains to be understood. Second, the estimated protoplanetary masses are certainly lower limits. Accretion from a surrounding massive circumplanetary disk or envelope would increase the final protoplanetary mass. The efficiency of protoplanetary accretion depends on many factors, which are beyond the scope of this study. Considering that tidal stripping of the inward-migrating clumps is rather efficient (see Figs.~\ref{fig:3} and \ref{fig:7}), we may expect that most of the circumplanetary disk and envelope will be lost and the resulting protoplanet will remain in the planetary mass regime. 
Taken into account that clumps are also overabundant in dust at their finals stages of evolution (see Table~\ref{tab:3})), metal-rich gas giants may form, which is in agreement with the recent analysis of mass-metallicity relation for giant planets \citep{2016Thorngren}.
Third, the destruction and regeneration of dusty rings discussed in the context inward-migrating clumps (see Figure~\ref{fig:ring}) may be affected by the formation of protoplanets at several tens of au. This issue needs further investigation with modified numerical simulations that take the formation of protoplanets into account. 
Finally, we have not considered dust back reaction on gas in our simulations. Although dust-to-gas ratio rarely exceeds 0.1 in the disk, this situation may change in higher resolution simulations and dust back reaction will be considered in our follow-up studies using the method laid out in \citet{2018Stoyanovskaya}.

\begin{table*}
\center
\caption{\label{tab:2} Characteristics of protoplanetary seeds}
\begin{tabular}{cccc|ccc}
\hline 
&  & model 1 & & & model 2 & \tabularnewline
\hline 
& $R_{\rm p.p.}=5 R_{\rm Jup}$ & $R_{\rm p.p.}=2.5 R_{\rm Jup}$ & $R_{\rm p.p.}=10 R_{\rm Jup}$ & $R_{\rm p.p.}=5 R_{\rm Jup}$ & $R_{\rm p.p.}=2.5 R_{\rm Jup}$ & $R_{\rm p.p.}=10 R_{\rm Jup}$\tabularnewline
\hline 
\hline 
Total mass [$M_{\rm Jup}$]  & 0.91 & 0.5 & 1.62 & 0.48 & 0.25 & 0.87\tabularnewline
Dust mass [$M_{\oplus}$]  & 3.1 & 1.7 & 5.5 & 1.9 & 1.0 & 3.5\tabularnewline
Dust-to-gas ratio & 0.011 & 0.011 & 0.011 & 0.013& 0.013 & 0.013\tabularnewline
\hline 
\end{tabular}
\end{table*}

\section{Conclusions}
\label{conclusions}
We studied the long-term evolution of strongly gravitationally unstable circumstellar disks using the numerical hydrodynamics code FEOSAD, which computes the joint dynamics of gas and dust (including dust growth) in the thin-disk limit \citep[see][for detail]{2018VorobyovAkimkin}. 
The simulations covered about 600~kyr of disk evolution starting from the instance of its formation through the gravitational collapse of a rotating pre-stellar core. Two model cores with slightly different initial parameters were considered.
Particular attention was paid to the evolution of dust, which consists of  two components: small sub-micron dust and grown dust with a varying maximum size. Thanks to the sub-au resolution, dust dynamics and growth inside gaseous clumps formed in the disk through gravitational fragmentation was studied in detail. 
Our main conclusions can be summarized as follows.

- The evolution of a gravitationally fragmenting disk has a highly time-varying character. Its nonaxisymmetric structures, such as spiral arms, rings, and clumps,  constantly  form,  evolve,  and  decay, meaning that such disks cannot be described by steady-state models.  Concurrently, the total dust-to-gas mass ratio becomes highly non-homogeneous throughout the disk extent and can strongly deviate from the initial canonical value of 1:100 to both lower and higher values.  

- The clumps formed through gravitational fragmentation have a complicated velocity pattern, which differs notably from the Keplerian rotation, particularly in the clump interiors. The radial gas surface density distribution can be approximated by a central plateau and a declining power-law tail. 

-  At the instance of clump formation, small sub-micron dust dominates throughout the clump extent. As the clumps evolve and migrate towards the star, small dust is efficiently converted into grown dust, reaching a maximum radius of several decimeters. Concurrently, grown dust drifts towards the clump center and forms a central core, which is notably more compact than the gaseous component of the clump.

- During their migration towards the star, the clumps disperse through the action of tidal torques. Before tidal dispersal, however, the gas temperature in the clump interiors can exceed 2000~K, which can lead to molecular hydrogen dissociation and second collapse, a process that is not captured in our numerical models because of still limited numerical resolution. We therefore foresee the formation of protoplanets at orbital distances of several tens of au with initial masses of gas and dust in the protoplanetary seeds in the (0.25--1.6)~$M_{\rm Jup}$ and (1.0--5.5)~$M_\oplus$ limits, respectively. The resulting protoplanetary seeds are slightly overabundant in dust as compared to the canonical 1:100 value. 

- Tidally disrupted clumps release processed dust at several tens of au from the star, leading to the formation of dusty rings at these distances. These rings may have a connection to ring-like structures at several tens of au found by the ALMA in youngest and massive protoplanetary disks \citep{2019vanderMarel}.
The lifetime of these rings is however limited by migration timescales of more distant clumps, which 
disturb these dusty structures as they migrate inward, forcing part of the dust to accrete through the central sink cell while redistributing the rest on more distant orbits in the disk.

 - Reasonable (tens of per cent) variations in the initial properties of collapsing pre-stellar cores (such as their mass, amount of rotation, initial temperature) do not influence our main conclusions on the dust migration and growth in gravitationally fragmenting disks, meaning that our results are robust.

We note that the final masses of gas and dust in the forming protoplanets depend on the efficiency of accretion and dispersal of material in massive circumlanetary disks/envelopes, which are likely to form around the protoplanets after the second collapse of the clumps. Taken into account that clumps are overabundant in dust at their finals stages of evolution, metal-rich gas giants may ultimately form,  in agreement with the recent analysis of mass-metallicity relation for giant planets \citep{2016Thorngren}.
We plan to address this issue in our subsequent studies employing a sub-grid model for the forming protoplanets.

\section*{Acknowledgements}
We are thankful to the referee for useful comments that helped to improve the manuscript.
This work was supported by the Russian Science Foundation grant 17-12-01168. V.G.E. acknowledges Swedish Institute for a visitor grant allowing to visit the Lund University. The simulations were performed on the Vienna Scientific Cluster (VSC-3).


\end{document}